\documentclass[11pt,a4paper]{article}                                           
\usepackage{cite}
\usepackage{mcite}
\usepackage{epsfig}

\epsfverbosetrue

\def\3{\ss}

\newcommand{\gev}{{\rm Ge}\kern-1.pt{\rm V}}
\newcommand{\gevsq}{\mbox{$\mathrm{{\rm Ge}\kern-1.pt{\rm V}}^2$}}
\newcommand{\gevmsq}{\mbox{$\mathrm{{\rm Ge}\kern-1.pt{\rm V}}^{-2}$}}
\newcommand{\sgeq}
{\mbox{\raisebox{-0.4ex}{$\;\stackrel{>}{\scriptstyle 
\sim}\;$}}}

\begin{document}
                   
\title{{\tt\normalsize\hspace*{28pt}DESY 00-084\hfill ISSN
0418-9833\hspace*{28pt} \\[-2ex]\hspace*{28pt}June 2000\hfill\
}\vskip2cm \bf\LARGE 
Measurement of exclusive {\boldmath $\omega$} electroproduction at HERA\\
\vskip2cm }               
                    
\author{ZEUS Collaboration}
\date{}
 
\maketitle
\begin{abstract}
\noindent

The exclusive electroproduction of $\omega$ mesons,
$ep \rightarrow e\omega p$, has been studied in the kinematic
range $3<Q^2<20$~{\gevsq}, $40 < W < 120$~{\gev} and $|t| < 0.6$~{\gevsq}
with the ZEUS detector at HERA using an integrated luminosity 
of 37.7 pb$^{-1}$.  
The $\omega$ mesons were identified via the decay $\omega \rightarrow
\pi^+\pi^-\pi^0$.  
The exclusive cross section in the above kinematic region is
$\sigma_{ep \rightarrow e\omega p} = 0.108 \pm 0.014(stat.) \pm
0.026(syst.)$~nb.
The reaction $ep \rightarrow e\phi p$, $\phi \rightarrow
\pi^+\pi^-\pi^0$, has also been measured. The cross sections, as well
as the ratios $\sigma_{\gamma^{*}p \rightarrow \omega
  p}/\sigma_{\gamma^{*}p \rightarrow \rho^0 p}$ and
$\sigma_{\gamma^{*}p \rightarrow \omega p}/\sigma_{\gamma^{*}p
  \rightarrow \phi p}$, are presented as a function of $W$ and $Q^2$.
Thus, for the first time, the properties 
of $\omega$ electroproduction can be compared to those
of $\rho^0,  \phi$ and $J/\psi$ electroproduction at high $W$.

\end{abstract}

\setcounter{page}{0}
\thispagestyle{empty}
\newpage
\pagenumbering{arabic}
\thispagestyle{empty} \newpage

%
%
%
%
\newcommand{\address}{ }                                                                           
\pagenumbering{Roman}                                                                              
\begin{center}                                                                                     
{                      \Large  The ZEUS Collaboration              }                               
\end{center}                                                                                       
  J.~Breitweg,                                                                                     
  S.~Chekanov,                                                                                     
  M.~Derrick,                                                                                      
  D.~Krakauer,                                                                                     
  S.~Magill,                                                                                       
  B.~Musgrave,                                                                                     
  A.~Pellegrino,                                                                                   
  J.~Repond,                                                                                       
  R.~Stanek,                                                                                       
  R.~Yoshida\\                                                                                     
 {\it Argonne National Laboratory, Argonne, IL, USA}~$^{p}$
% ...
\newpage                                                                                           
$^{\    1}$ now visiting scientist at DESY \\                                                      
% ...
                                                           %
                                                           %
\newpage   
                                                           %
                                                           %
\begin{tabular}[h]{rp{14cm}}                                                                       
$^{a}$ &  supported by the Natural Sciences and Engineering Research                               
          Council of Canada (NSERC)  \\                                                            
$^{b}$ &  supported by the FCAR of Qu\'ebec, Canada  \\                                            
$^{c}$ &  supported by the German Federal Ministry for Education and                               
          Science, Research and Technology (BMBF), under contract                                  
          numbers 057BN19P, 057FR19P, 057HH19P, 057HH29P, 057SI75I \\                              
$^{d}$ &  supported by the MINERVA Gesellschaft f\"ur Forschung GmbH, the                          
German Israeli Foundation, the Israel Science Foundation, the Israel                               
Ministry of Science and the Benozyio Center for High Energy Physics\\                              
$^{e}$ &  supported by the German-Israeli Foundation, the Israel Science                           
          Foundation, the U.S.-Israel Binational Science Foundation, and by                        
          the Israel Ministry of Science \\                                                        
$^{f}$ &  supported by the Italian National Institute for Nuclear Physics                          
          (INFN) \\                                                                                
$^{g}$ &  supported by the Japanese Ministry of Education, Science and                             
          Culture (the Monbusho) and its grants for Scientific Research \\                         
$^{h}$ &  supported by the Korean Ministry of Education and Korea Science                          
          and Engineering Foundation  \\                                                           
$^{i}$ &  supported by the Netherlands Foundation for Research on                                  
          Matter (FOM) \\                                                                          
$^{j}$ &  supported by the Polish State Committee for Scientific Research,                         
          grant No. 112/E-356/SPUB/DESY/P03/DZ 3/99, 620/E-77/SPUB/DESY/P-03/                      
          DZ 1/99, 2P03B03216, 2P03B04616, 2P03B03517, and by the German                           
          Federal Ministry of Education and Science, Research and Technology (BMBF)\\              
$^{k}$ &  supported by the Polish State Committee for Scientific                                   
          Research (grant No. 2P03B08614 and 2P03B06116) \\                                        
$^{l}$ &  partially supported by the German Federal Ministry for                                   
          Education and Science, Research and Technology (BMBF)  \\                                
$^{m}$ &  supported by the Fund for Fundamental Research of Russian Ministry                       
          for Science and Edu\-cation and by the German Federal Ministry for                       
          Education and Science, Research and Technology (BMBF) \\                                 
$^{n}$ &  supported by the Spanish Ministry of Education                                           
          and Science through funds provided by CICYT \\                                           
$^{o}$ &  supported by the Particle Physics and                                                    
          Astronomy Research Council \\                                                            
$^{p}$ &  supported by the US Department of Energy \\                                              
$^{q}$ &  supported by the US National Science Foundation                                          
\end{tabular}                                                                                      
                                                           %
                                                           %
\newpage
\parindent0.6cm \parskip0.0cm
\pagenumbering{arabic}

\section{Introduction}


The exclusive photoproduction of vector mesons, $\rho^0$, $\omega$,
$\phi$ and $J/\psi$, has been studied over a wide range of the
photon-proton centre-of-mass energy $W$~\cite{
rmp_50_261,
special1,
special2,
epj_2_247,
zfp_73_73,
pl_377_259,
zfp_75_215,
dr_00_037}.  
For high $W$ and for light vector mesons, these
reactions display features characteristic of soft diffractive processes,
and
are well described
within the framework of the vector dominance model (VDM)
\cite{prl_22_981} and Regge phenomenology \cite{collins}.
In the VDM, the photon is assumed to fluctuate into a vector meson (VM), which
subsequently interacts with the target proton.  
The expectation of the VDM and Regge phenomenology 
is that the cross sections for 
exclusive VM production will be in proportions determined by 
the couplings of the photon to the vector mesons and by the 
elastic VM-proton cross sections. 
The couplings, in particular, 
are determined by the quark current decomposition of the photon 
and by the quark wave function of the VM. 
The SU(4)
prediction, which
ignores the VM mass differences,
is that the coupling strengths of the photon to the 
$\rho^0, \omega, \phi$ and $J/\psi$ mesons are in the 
ratio\footnote{If the coupling strengths are calculated from
the VM masses and the measured partial widths, $\Gamma_{V \rightarrow
e^+e^-}$, the ratio becomes 9:0.8:2.4:28} 9:1:2:8.
Both at fixed target experiments at $W \sgeq 8$~GeV~\cite{rmp_50_261} 
and at HERA~\cite{zfp_73_73,pl_377_259,zfp_75_215}, 
the ratio $\omega / \rho^0$ of the exclusive
photoproduction cross sections is found to be approximately~1:9, 
while the $\phi / \rho^0$ and $(J/\psi) / \rho^0$ ratios 
are found to be smaller than 
the SU(4) predicted ratios of photon-VM couplings.


For high-energy exclusive photoproduction of heavy VMs and for
electroproduction of all VMs at large virtualities, $Q^2$, of the
exchanged virtual photon, $\gamma^*$, an alternative production
mechanism has been proposed~\cite{pr_50_3134,pr_54_3194,koepf}: the
virtual photon fluctuates into a $q\bar{q}$ pair before arriving
at the target, and it is this $q\bar{q}$ state that scatters elastically
off the proton.  The VM is formed well after the
interaction.  If the transverse size of the $q\bar{q}$ fluctuation is
small enough, the interaction is expected to become flavour independent
and the gluons of the proton are resolved~\cite{Collins:1997fb}.  
Such a configuration occurs when the mass of the quarks is large 
or if the photon with large virtuality is longitudinally polarized.  
In these cases, perturbative 
QCD can be 
applied~\cite{pr_50_3134,pr_54_3194,Ryskin:1993ui,Ryskin:1995hz,pr_55_4329}
and the expectation for the ratios of VM production
cross sections is given by the  ratios of the couplings. 

The HERA measurements of exclusive electroproduction of 
$\rho^0$~\cite{epj_6_603,epj_13_371}, 
$\phi$~\cite{pl_380_220,dr_00_070} and
$J/\psi$~\cite{epj_6_603,epj_10_373} mesons,
as well as the exclusive photoproduction of 
$J/\psi$~\cite{zfp_75_215,dr_00_037} and
$\Upsilon$~\cite{dr_00_037,pl_437_432} mesons, 
are in broad agreement with the expectations from QCD. 
In particular, the sharp decrease of the cross sections with
$Q^2$, the strong rise of the
$J/\psi$ cross section with $W$, the change in the $W$ dependence and
the broadening of the $t$ distribution of the $\rho^0$ cross section
with increasing $Q^2$ show the expected behaviour.
With increasing $Q^2$, the $\phi / \rho^0$ and $(J/\psi) / \rho^0$ cross-section 
ratios rise towards the expected values.

This letter reports the
first measurement of
the exclusive $\omega$ electroproduction cross section for
$40<W<120$~{\gev} and $3<Q^2<20$~{\gevsq}.
The $\omega$ mesons were identified via the decay $\omega \rightarrow
\pi^+\pi^-\pi^0$. 
Measurements of the exclusive $\phi$ electroproduction 
cross section using the $\pi^+\pi^-\pi^0$ final state are also reported.
The compatibility of the $\omega / \rho^0$ and 
$\omega / \phi$ cross-section ratios with the SU(4) hypothesis is investigated.
Thus, the data presented here permits, for the first time,
a discussion of the ratios of the production cross sections
of all VMs related by SU(4).

\section{Experimental conditions}


The measurements were performed at the $ep$ collider HERA with
the ZEUS detector
using an integrated luminosity of $37.7~{\rm pb^{-1}}$.  
During 1996 and 1997 HERA operated with a proton
energy of 820~{\gev} and a positron energy of 27.5~{\gev}.  A detailed
description of the ZEUS detector can be found
elsewhere~\cite{zeusdetector}.  The main components used in this
analysis are described below.


The high-resolution uranium-scintillator calorimeter
CAL~\cite{special3} consists of three
parts: forward\footnote{Throughout this paper the standard ZEUS
  right-handed coordinate system is used: the $Z$-axis points in the
  direction of the proton beam momentum (referred to as the forward
  direction) and the horizontal $X$-axis points towards the centre of
  HERA. The nominal interaction point is at $X=Y=Z=0$.
  The polar angle $\theta$ is defined with respect to the 
  positive $Z$-direction.} (FCAL),
barrel (BCAL) and rear (RCAL) calorimeters.  Each part is subdivided
transversely into towers, which are segmented longitudinally into one
electromagnetic section (EMC) and one (RCAL) or two (FCAL, BCAL)
hadronic sections (HAC).  
Each section is further subdivided into cells, which are the units that are read out.
The relative energy resolutions of the calorimeter,
as determined in a test beam, are $\sigma_{E}/E = 0.18/\sqrt{E}$ for
electrons and $\sigma_{E}/E = 0.35/\sqrt{E}$ for hadrons, where $E$ is
expressed in~{\gev}.


Charged-particle tracks  are reconstructed and their momenta determined
using  the central tracking detector 
(CTD)~\cite{special4}. 
The CTD is a cylindrical drift chamber operated 
in a magnetic field of 1.43 T
produced by a thin superconducting solenoid.
The CTD consists of 72 cylindrical layers, organised in 9 superlayers
covering the polar angular region $15^\circ  < \theta < 164^\circ $.
The transverse momentum
resolution for full-length tracks is $ \sigma _{p_\perp }/p_\perp
=0.0058\, p_\perp \oplus 0.0065\oplus {0.0014/ p_\perp }$ $( p_\perp$ in GeV).


The positions of positrons scattered at small angles with respect to
the beam direction are determined by the small-angle rear tracking
detector (SRTD). The~SRTD~is attached to the front face of the RCAL.
It consists of two planes of scintillator strips, 1 cm wide and 0.5 cm
thick, arranged in orthogonal orientations and read out via optical
fibres and photomultiplier tubes. It covers the region of $68\times68$
cm$^2$ in $X$ and $Y$, except for a $10\times20$ cm$^2$ hole
at the centre for the beam pipe.  The SRTD has a position resolution
of 0.3 cm.


The luminosity is determined from the rate of the Bethe-Heitler
process $ e^+ p \rightarrow e^+ \gamma p$, where the high-energy
photon is detected in a lead-scintillator calorimeter located
at $Z= - 107$~m in the HERA tunnel downstream of the interaction point
in the positron flight direction~\cite{special5}.

\section{Kinematics and cross sections}

\begin{figure}[p]
\centering
\leavevmode
\epsfxsize=10.0cm\epsfbox{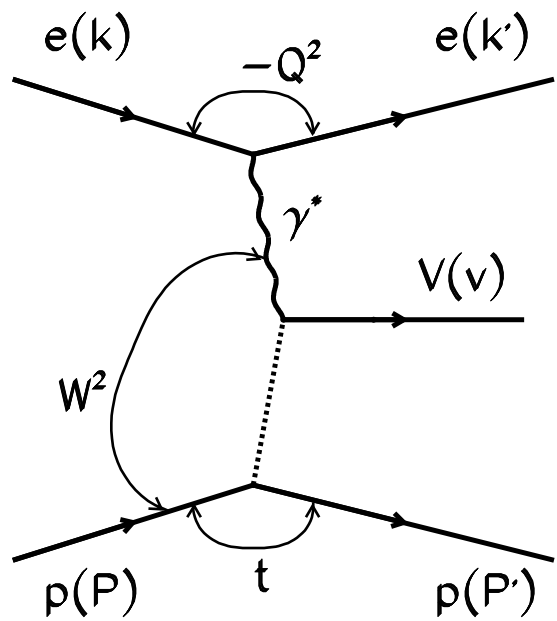}
\caption{
Schematic diagram of exclusive vector-meson
electroproduction in $e^+p$ interactions, $ep \rightarrow eVp$. 
In this analysis, $V$ is either an $\omega$ or $\phi$ meson.}
\label{fig:graph}
\end{figure}

Figure~\ref{fig:graph} shows a schematic diagram for the reaction:  
\[
e(k)~p(P) ~\rightarrow ~e(k')~V(v)~p(P')
\]
where $V$ is an $\omega$ or $\phi$ meson and  
$k$, $k'$, $P$, $P'$, and $v$ are the four-momenta of the incident
positron, scattered positron, incident proton, scattered proton and
vector meson, respectively.  
The kinematic variables used to describe exclusive VM
production are:
\begin{itemize}
\item $Q^2 = -q^2=-(k-k')^2$, the negative squared four-momentum  
of the virtual photon;
\item $W^2 = (q+P)^2 $, the squared centre-of-mass energy of the
  photon-proton system;
\item $y = (P \cdot q)/(P \cdot k)$, the fraction of the positron
  energy transferred to the photon in the proton rest frame;
\item $t = (P - P')^2 = (v - q) ^2$, the squared
four-momentum-transfer at the proton vertex.
\end{itemize}


The kinematic variables were reconstructed using the
``constrained'' method. This method uses the momenta of the decay
charged particles measured in the CTD, the momenta of the two photons 
from the $\pi^0 \rightarrow \gamma \gamma$ decay, 
measured as described in Section 5, and the polar and
azimuthal angles of the scattered positron measured by the CAL and the SRTD.
Neglecting the transverse momentum of the outgoing proton with respect
to its incoming momentum, the energy of the scattered positron can be
expressed as:
\[
E_{{e}^{\prime}} \simeq 
[2E_{e} -(E_{V} - p^Z_{V})]/(1-\cos{\theta_{{e}^{\prime}}})
\]
where $E_{e}$ is the energy of the incident positron, $E_{V}$
and $p^Z_{V}$ are the energy and longitudinal momentum of the
vector meson $V$, and $\theta_{{e}^{\prime}}$ is the polar angle of the
scattered positron.  The value of $Q^2$ was calculated from:
\[
Q^2 = 2E_{{e}^{\prime}} E_{{e}} (1 + \cos{\theta_{{e}^{\prime}}})
\]
and $W$ and $t$ were calculated as described above.


In the Born approximation,
the virtual photon-proton cross section, $\sigma^{\gamma{^*} p}$,
can be determined from the measured positron-proton cross section:
\[ 
\sigma^{\gamma^{*} p} =
\sigma_{\rm T}^{\gamma^{*} p}+ \ \epsilon \sigma_{\rm
  L}^{\gamma^{*} p} = \frac{1}{{\it \Gamma}_{\rm T}(Q^2,y)}
\cdot \frac{d^2 \sigma^{ep}}{dQ^2 dy}
\] 
where $\epsilon=2(1-y)/[1+(1-y)^2]$,
${\it \Gamma}_{\rm T}$ is the flux of transverse photons
and $\sigma_{\rm T}^{\gamma^{*} p}$ and
$\sigma_{\rm L}^{\gamma^{*} p}$ are the transverse and the 
longitudinal virtual photoproduction cross sections, respectively.
The cross-section $\sigma^{\gamma^{*} p}$
is used to evaluate the total exclusive
cross section, $\sigma_{\rm tot}^{\gamma^{*} p} =
\sigma_{\rm T}^{\gamma^{*} p}+ \ \sigma_{\rm
  L}^{\gamma^{*} p}$, through the relation:
\[
\sigma_{\rm tot}^{\gamma^{*} p} = \
\frac{1+R}{1+\epsilon R} \sigma^{\gamma^{*} p}
\]
where $R=\sigma_{\rm L}^{\gamma^{*} p} / \sigma_{\rm
  T}^{\gamma^{*} p}$.
In the kinematic range of this measurement, the value of $\epsilon$ is close 
to unity, and  $\sigma^{\gamma^{*} p}$ differs 
from  $\sigma_{\rm tot}^{\gamma^{*} p}$ by less than one percent.

\section{Event selection}
Events were selected online with a three-level trigger system. 
Offline, the following requirements
were imposed to select candidates for the reaction $e^+p \rightarrow
e^+\pi^+\pi^-\pi^0p$:
\begin{itemize}
\item the energy of the scattered positron, measured in the CAL, 
  was required to be greater
  than 10~{\gev};
\item the interaction vertex was required to have its $Z$ coordinate
  within $\pm 50$ cm of the nominal interaction point and to lie within  
  a transverse distance of 0.6 cm of the nominal beam position;
\item in addition to a scattered positron, two oppositely
  charged tracks were required, each associated with the reconstructed
  vertex, and each with pseudorapidity\footnote{The pseudorapidity 
  $\eta$ is defined as $\eta=-\ln$[tan$\frac{\theta}{2}$].}
 $|\eta |< 1.75$ and transverse momentum greater than 150~MeV;
\item in addition to the energy deposits in the calorimeter matched to
the scattered positron and to
  the above-mentioned tracks, two further electromagnetic energy
  deposits in the calorimeter were required (see Section 5); 
\item to suppress the contribution from proton dissociation
(see Section 7), events with 
  energy deposits above 800 MeV in the forward
  calorimeter within a radius of 50 cm from the $Z$ axis
  were rejected;
\item to reduce the corrections from initial-state radiation, a cut on the
  difference in total energy and total longitudinal momentum, as measured
  in the CAL, of $E-p_Z > 40$~{\gev} was applied. 
\end{itemize}

In addition, the following cuts were applied to select a kinematic
region of high acceptance: $3 < Q^2 <20$~{\gevsq}, $40 < W <
120$~{\gev} and $|t| < 0.6$~{\gevsq}.  Only events in the $\pi^+\pi^-$
mass interval $0.3 < M_{\pi \pi} < 0.6$~{\gev} were kept for further
analysis.

\section{Reconstruction of the {\boldmath $\pi^0$}}

The two-photon decay of the $\pi^0$ was reconstructed using signals
in the calorimeter combined into condensates, which are objects
consisting of adjacent calorimeter cells. 
To reject background from uranium radioactivity in the calorimeter
cells, a minimum individual cell energy of 100 MeV was required.
Only condensates consisting solely 
of cells in the EMC section of the CAL were used.
The position and energy, $E_\gamma$, of these condensates were then
used to determine the invariant mass, $M_{\gamma \gamma}$, of the photon
candidates, assuming the event vertex as measured by the CTD.

\begin{figure}[htb]
\centering
\leavevmode
\epsfxsize=16.0cm\epsfbox{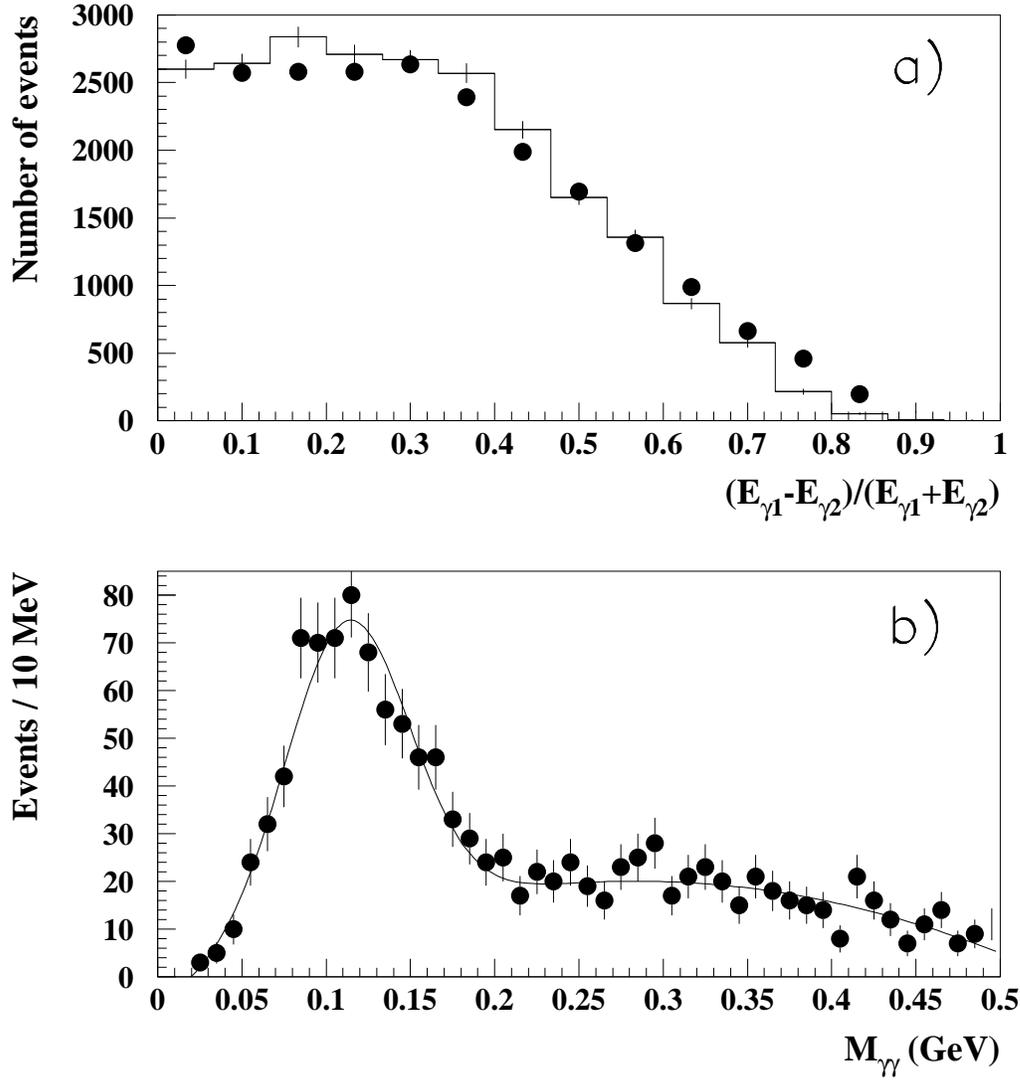}



\caption{
a) The ratio $(E_{\gamma 1} - E_{\gamma 2})/(E_{\gamma 1} + 
E_{\gamma 2})$ for $\pi^0$ candidates (solid points), 
where $\gamma 1$ is the more energetic photon
candidate.
The histogram corresponds to the MC simulation described in the text.
The vertical bars on the histogram correspond to the 
statistical uncertainties of the MC. 
The statistical uncertainties of the data are smaller than the size of the
solid points.
b) The invariant-mass spectrum for two photons (solid points). 
The line represents a fit to the sum of 
a Gaussian and a second-order polynomial as described in the text.
}
\label{fig:e1e2}
\end{figure}

Uranium noise and CAL cells not correctly matched to charged tracks 
can yield condensates which are misidentified as low-energy photons.
Such condensates introduce a background contribution to
the $M_{\gamma \gamma}$ spectrum and were largely removed by requiring
a minimum photon energy of 200 MeV and  
$E_{\gamma 1}+E_{\gamma 2} > 600$ MeV,
where $\gamma 1$ is the more energetic photon.
These energy
requirements reduced the acceptance of the analysis by a factor of two, but
reduced the systematic
uncertainty on the background subtraction by a factor of three.
The quantity 
${|E_{\gamma 1} - E_{\gamma 2}|}/ {(E_{\gamma 1} + E_{\gamma 2})}$,
which is particularly sensitive to backgrounds, 
is shown in  Fig.~\ref{fig:e1e2}a.
The data and the Monte Carlo (MC) simulation described in Section 6
agree well after the above requirements are applied.
The resulting $M_{\gamma \gamma}$ spectrum is shown in
Fig.~\ref{fig:e1e2}b, where a clear $\pi^0$ mass peak is seen.  A fit to the
sum of a Gaussian function and a second-order polynomial yielded 
a mean value of the Gaussian of $110 \pm 2 $ MeV, in good agreement with 
simulation studies of the detector response.
The difference
with respect to the nominal $\pi^0$ mass is mainly due to the
energy loss in the superconducting solenoid in front of the calorimeter
and was implicitly corrected for by the fit explained below.
Only events with $M_{\gamma \gamma} < 200$~MeV were 
retained for further analysis.

To improve the resolution in the momentum of the $\pi^+ \pi^- \pi^0$
system, the invariant mass of the two photons was constrained to the
$\pi^0$ mass.
The angle $\alpha$ between
the two photons is well determined by the 
position information of the condensates,
so that the resolution in $M_{\gamma\gamma}$ is dominated
by the energy resolution.  Thus, only the energies of the photons
were varied in this procedure. The modified values $E_{\gamma 1}^{\rm fit}$,
$E_{\gamma 2}^{\rm fit}$ of the energies $E_{\gamma 1}$, 
$E_{\gamma 2}$ of the CAL condensates were
determined by minimising the quantity:
\[
\chi^2(E_{\gamma 1}^{\rm fit},E_{\gamma 2}^{\rm fit}) = 
\frac{(E_{\gamma 1} - E_{\gamma 1}^{\rm fit})^2}{\sigma_{E_{\gamma 1}^{\rm fit}}^2}+
\frac{(E_{\gamma 2} - E_{\gamma 2}^{\rm fit})^2}{\sigma_{E_{\gamma 2}^{\rm fit}}^2}\, 
\]
using the constraint:
$$
M_{\pi^0}= \sqrt{2\cdot E_{\gamma 1}^{\rm fit}E_{\gamma 2}^{\rm fit}\cdot (1-\cos
  \alpha)} \, $$
where $\sigma_{E_{\gamma i}^{\rm fit}}({\gev})\propto
\sqrt{E_{\gamma i}^{\rm fit}({\gev})}$ are the corresponding energy resolutions
of the calorimeter.


\section{Monte Carlo simulation and acceptance corrections}

A MC generator for exclusive electroproduction of 
light VMs~\cite{muchorthesis},
interfaced to HERACLES~\cite{heracles} to simulate
radiative effects, was used to evaluate the acceptance. 
In this generator the cross sections were parameterised 
over the entire $W$ and $Q^2$ range using the ZEUS data 
on $\rho^0$ electroproduction~\cite{epj_6_603} in terms of
$\sigma^{\gamma^* p \rightarrow V p}_{\rm L}$ and 
$\sigma^{\gamma^* p \rightarrow V p}_{\rm T}$,
where $V = \rho^0, \omega$~or $\phi$; $s$-channel helicity
conservation was assumed.
The MC samples generated in this way display good agreement
with the present data, for both $\omega$ and $\phi$ production.
Whenever the results of the current analysis are compared to previously 
published cross sections at slightly different 
values of $Q^2$ and $W$, the cross sections have been translated to 
appropriate values using the ZEUS $\rho^0$ cross-section
parameterisation.

The average acceptance for $\omega$ mesons
is 2.5$\%$, an order of magnitude lower than that for the
corresponding ZEUS $\rho^0$ analysis.  This is
mainly caused by the low $\pi^0$ reconstruction efficiency.
The corresponding acceptance for $\phi$ mesons is 2.3\%.

The Monte Carlo program DIPSI~\cite{cpc_100_195}, based on the model of
Ryskin~\cite{Ryskin:1993ui}, was used for systematic checks.


\section{Background}
The main source of resonant background to the exclusive reaction
$e p \rightarrow e V p$  is the
proton-dissociative reaction 
$e p \rightarrow e V N$, where $N$ is a 
hadronic system produced by the dissociation of the proton. 
The proton-dissociative events in which the
hadronic system $N$ deposits energy around the beam pipe in the FCAL 
are removed by the selection criteria;
the rest are misidentified as 
$e p \rightarrow e V p$ reactions and therefore have to be subtracted.
The proton-dissociative fraction of events was estimated in the 
ZEUS $\rho^0$
analysis to be $(24^{+9}_{-5}) \%$ for $|t| < 0.6$~{\gevsq},
independent of $Q^2$ and $W$.  
The same fraction was assumed in this study.


\section{Results}

\subsection{Analysis of the mass spectrum \label{sec:masssp}}

The invariant-mass ($M_{3\pi}$) spectrum for the $\pi^+\pi^-\pi^0$
system after all offline cuts is shown in Fig.~\ref{fig:mass}a.  In
addition to the $\omega$ signal, a second peak is visible from the process
$e p\rightarrow e \phi p$ (\mbox{$\phi\rightarrow\pi^+\pi^-\pi^0$}).  
The spectrum was fitted with the function:
\[ 
f(M_{3\pi}) = g_1(M_{3\pi}) + 
 g_2(M_{3\pi}) + \zeta(M_{3\pi})\, 
\]
where $g_1$ and $g_2$ are Gaussian functions to describe the
$\omega$ and $\phi$ mass peaks, and $\zeta$ is a second-order
polynomial representing the background. The latter is mainly due to
the background under the $\pi^0$ peak, as well as to other
processes in which an $e \pi^+ \pi^- \pi^0$ final state is detected.
The fitted values of the $\omega$ and $\phi$ masses are $787\pm 5$ MeV and
$1019\pm 10$ MeV, respectively, compatible with their nominal
values~\cite{epj_3_1} and with MC expectations. 
The values obtained for 
the widths are dominated by the detector resolution.
The standard deviations of the Gaussian resolution functions 
are $38$ MeV for the $\omega$ 
and $36$ MeV for the $\phi$, in accord with the MC predictions.

The number of $\omega$ and $\phi$ candidates observed after background
subtraction was determined by integrating the corresponding Gaussian
function, which yielded $N_{\omega}=116 \pm 14$ and 
$N_{\phi}=38 \pm 11$.

\subsection{Cross sections and systematic uncertainties}

The cross sections for the reaction 
$ ep \rightarrow eVp $, where $V = \omega$ or $\phi$, 
were determined using the expression:
\[
\sigma_{ep \rightarrow eVp}  = \frac {N_V \cdot \Delta} 
{A_V \cdot L \cdot B_V}
\]
where $N_V$ is the total number of observed events,
 $L$ is the integrated luminosity,
$\Delta$ is the correction for the proton-dissociation background, and
$A_V$ and 
$B_V$ are the overall acceptance and branching ratio,
respectively. 

The systematic uncertainties are dominated by the uncertainty in
the extraction of the number of $\omega$ or $\phi$ events 
and the uncertainty of the proton-dissociative background.  
The extraction of the number of events
dominates the uncertainty because of the large number (nine) 
of free parameters needed to fit the mass distribution and
the small number of signal events.
When different
$M_{\pi\pi}$ or $M_{\gamma \gamma}$ cuts were applied, the shape of the
mass spectrum changed, which led to differences in the values of the
fit parameters, and hence the number of resonant events. 
The systematic uncertainty due to this procedure for extracting the
number of events is about $15\%$.  The uncertainty in the
proton-dissociative background was taken from the ZEUS $\rho^0$
analysis.

In the kinematic range
$3<Q^2<20$~{\gevsq}, $40<W<120$~{\gev} and $|t|<0.6$~{\gevsq},
the $\omega$ and $\phi$ exclusive electroproduction cross sections are
$\sigma_{ep \rightarrow e\omega p} = 0.108 \pm 0.014(stat.) \pm
0.026(syst.)$~nb and
$\sigma_{ep \rightarrow e\phi p} = 0.220 \pm 0.064(stat.) \pm
0.077(syst.)$~nb. 
%
The corresponding $\gamma^* p$ cross sections are presented 
in Table~\ref{tab:cross}.

\begin{figure}[htb]
\centering
\leavevmode
\epsfxsize=16.0cm\epsfbox{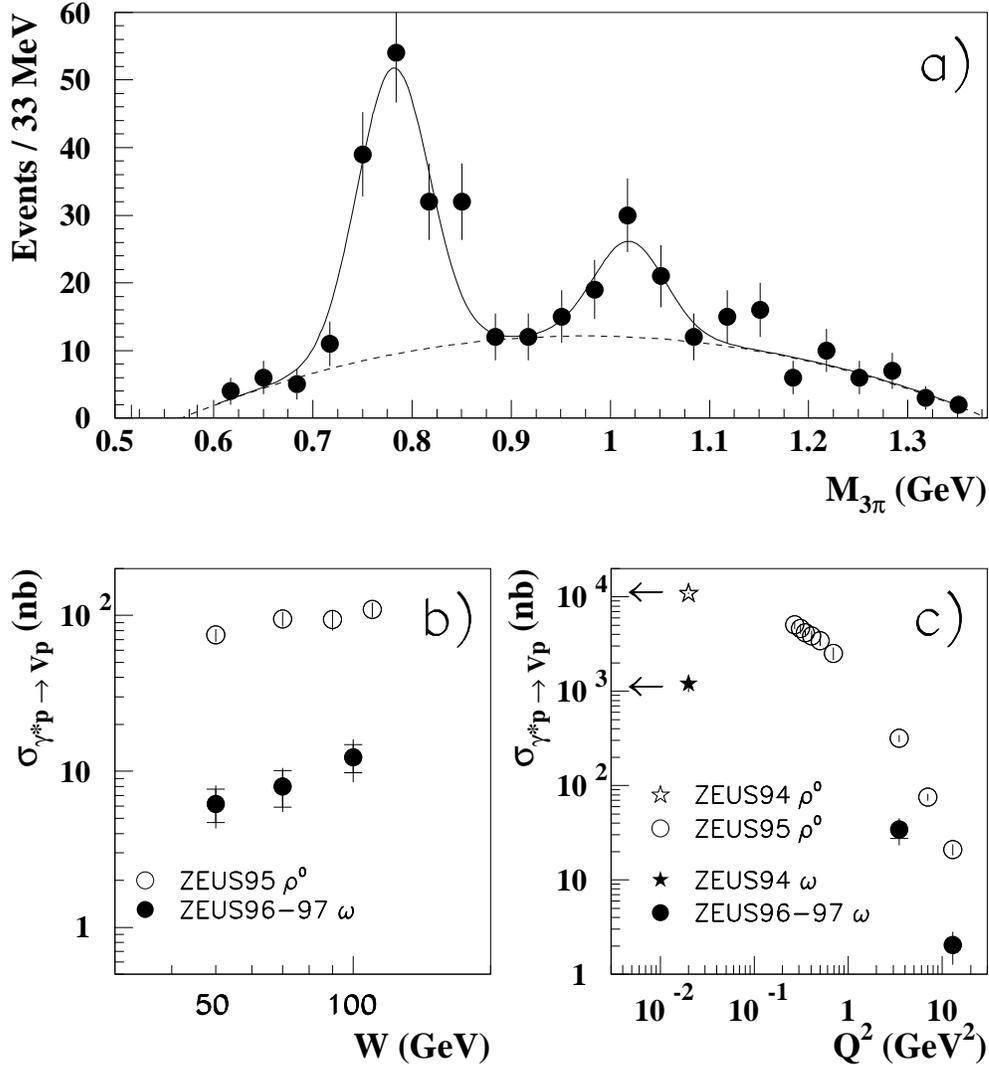}


\caption{a) 
The $\pi^+\pi^-\pi^0$ invariant-mass spectrum (solid points).
The two peaks 
correspond to the 
$\omega$ and $\phi$ mesons. The full line shows the 
result of the fit explained in the text. The dashed line shows the
fitted background.
b) The dependence of
$\sigma_{\gamma^*p \rightarrow \omega p}$ on $W$ at $Q^2$ 
= 7~{\gevsq}. The full circles are the results from this analysis, and
the open circles are from the ZEUS $\rho^0$ data
\protect\cite{epj_6_603}.
c) The dependence of  
$\sigma_{\gamma^*p \rightarrow \omega p}$ on $Q^2$ at $W$ = 70~{\gev}.
The results from this analysis are shown as full circles, while 
the full star is the photoproduction
result~\protect\cite{zfp_73_73}.
The open circles and star are ZEUS results on
the  $\rho^0$ cross sections at $W = 50$~{\gev} 
\protect\cite{epj_2_247,epj_6_603}. 
}
\label{fig:mass}
\end{figure}

The measured cross sections for exclusive $\omega$ production as a
function of $W$ and $Q^2$ are presented in 
Figs.~\ref{fig:mass}b and ~\ref{fig:mass}c.
The corresponding values are given in 
Tables~\ref{tab:crossW} and~\ref{tab:crossRatios}.
The exclusive $\omega$ photoproduction data 
($Q^2 \simeq 0$ GeV$^2$) displayed in Fig.~\ref{fig:mass}c 
were taken from a previous ZEUS publication~\cite{zfp_73_73}.  
The $\rho^0$ data~\cite{epj_2_247,epj_6_603}
are also shown for comparison.
The measured $Q^2$ and $W$ dependences for exclusive $\omega$ production
are consistent with those for $\rho^0$ meson production~\cite{epj_6_603}.
\subsection{Cross-section ratios}

The $\omega/\phi$, $\omega/\rho^0$, 
and $\phi/\rho^0$ cross-section ratios, 
obtained at $W=70$ GeV for the whole data sample
measured here, are listed in Table~\ref{tab:cross}.

The ratio $\omega/\phi = 0.49 \pm 0.15(stat.) \pm 0.12(syst.)$,
obtained with the present data, 
is consistent with the value expected from flavour independence.
In this case, most of the systematic uncertainties are common to
the $\omega$ and $\phi$ cross-section measurements 
and cancel in the ratio.

\begin{figure}[htb]

\leavevmode
\epsfxsize=16.0cm\epsfbox{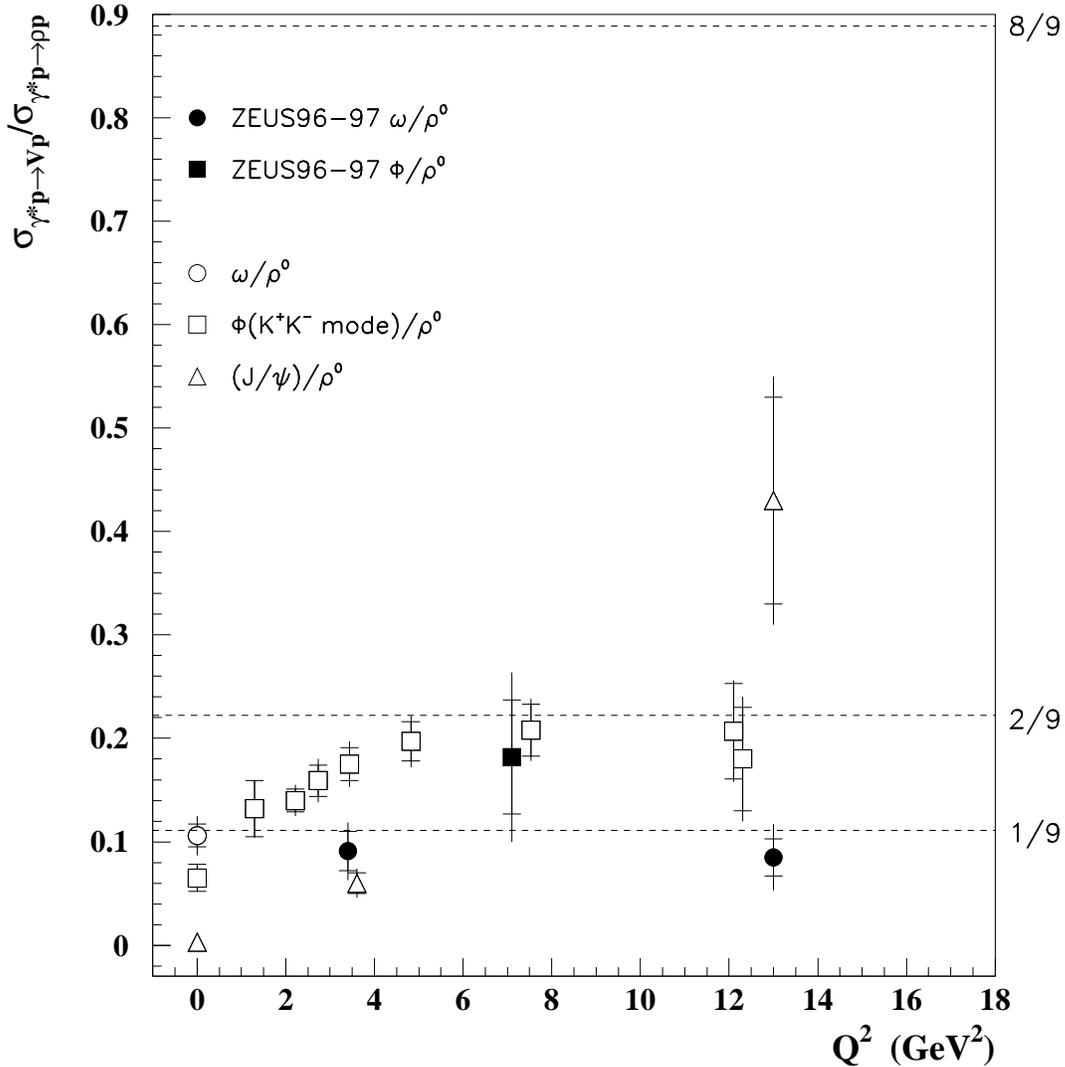}

\caption
{
The ratios of the $\omega$, $\phi$ and $J/\psi$ exclusive cross sections 
to those for the $\rho^0$ meson as a function of $Q^2$.
The $\omega/\rho^0$ and $\phi / \rho^0$ results from this analysis are 
shown with full symbols. The $\omega/\rho^0$ 
photoproduction point (open circle) 
was calculated using the measured cross sections from 
previous ZEUS publications
\protect\cite{zfp_73_73} and \protect\cite{epj_2_247}.
Open squares at $Q^2 \simeq 0$ GeV$^2$~\protect\cite{pl_377_259} 
and $Q^2= 12.3$ GeV$^2$~\protect\cite{pl_380_220}
represent the ZEUS  $\phi$/$\rho^0$ cross-section ratios,
while the others represent the H1 $\phi$/$\rho^0$ ratios
\protect\cite{dr_00_070};
for both the ZEUS and H1 measurements, 
the $\phi \rightarrow K^+K^-$ channel was used.
Triangles  represent the $(J/\psi)/\rho^0$ cross-section ratios
 \protect\cite{zfp_75_215,epj_6_603}.
The horizontal dashed lines correspond to the SU(4) expectations.
}
\label{fig:Xsec}
\end{figure}

In Table~\ref{tab:crossRatios} and Fig.~\ref{fig:Xsec},
the ratios of the exclusive $\omega$ to $\rho^0$ production cross
sections are shown for three $Q^2$ values. 
The ratio at $Q^2 \simeq 0~{\gevsq}$ was calculated using measured
$\omega$ \protect\cite{zfp_73_73} and $\rho^0$ \protect\cite{epj_2_247}
photoproduction cross sections at $W=80$ and $W=75$~{\gev}, respectively, 
whereas the ratios at higher $Q^2$ are for $W=70$~{\gev}. 
Due to the weak $W$ dependence in photoproduction, 
a direct comparison can be made without correction. 
No $Q^2$ dependence of the $\omega / \rho^0$ 
cross-section ratio is observed.
The ratio is consistent with the value expected
from flavour independence.


The $\phi/\rho^0$ cross-section ratio from Table~\ref{tab:cross}
is shown in Fig.~\ref{fig:Xsec}.
Also shown in Fig.~\ref{fig:Xsec} are the
H1 data at $Q^2 > 1$ GeV$^2$~\cite{dr_00_070}
and ZEUS data at $Q^2 \simeq 0$ GeV$^2$~\cite{pl_377_259} 
and $Q^2= 12.3$ GeV$^2$~\cite{pl_380_220},
all obtained using the $\phi \rightarrow K^+K^-$ channel. 
The present result is consistent with the earlier measurements.
The $\phi/\rho^0$ ratio approaches the SU(4) value for $Q^2 \sgeq 6$~GeV$^2$.

For both the $\omega / \rho^0$ and the $\phi/\rho^0$
cross-section ratios presented in this paper, 
the systematic uncertainties on the subtraction of the
proton-dissociative background cancel in the ratios. 
The remaining systematic uncertainties
were assumed to be dominated by the uncertainties on the 
$\omega$ and $\phi$ cross section measurements.

For completeness, the $(J/\psi) / \rho^0$ cross-section 
ratios\footnote{The recent H1 measurements on $\rho^0$ and $J/\psi$  
electroproduction~\cite{epj_13_371,epj_10_373} are not displayed 
in Fig.~\ref{fig:Xsec} 
since the $(J/\psi) / \rho^0$ cross-section ratios 
with systematic uncertainties were not published.}
at $W=90$ GeV~\cite{zfp_75_215,epj_6_603} 
are displayed in Fig.~\ref{fig:Xsec}.
A clear increase of the $(J/\psi) / \rho^0$ ratio with increasing $Q^2$
is observed, although the ratio is below the SU(4) value 
even at the largest $Q^2$.
A QCD-based model~\cite{pr_54_3194} is in agreement with this behaviour.

\section{Conclusions}

The reaction $ep \rightarrow e \omega p$ 
 has been studied in the $e \pi^+\pi^-\pi^0 p$ final state for 
$3<Q^2<20$~{\gevsq}, $40<W<120$~{\gev} and $|t| < 0.6$~{\gevsq}.
The cross-sections 
$\sigma_{ep \rightarrow
  e\omega p}$ and $\sigma_{\gamma^* p \rightarrow \omega p}$ have
been measured for the first time at high $Q^2$.  
In addition, the cross-sections
$\sigma_{ep \rightarrow  e\phi p}$ and
$\sigma_{\gamma^* p \rightarrow \phi p}$ have been measured
using the same final state. 

The cross-section ratio
$\omega/\phi = 0.49 \pm 0.15(stat.) \pm 0.12(syst.)$, 
obtained with the present data,
is consistent with the SU(4) expectation.

For the light vector mesons, $\rho^0, \omega$ and $\phi$, 
the cross-section ratios
are consistent with the SU(4) expectation for $Q^2 \sgeq 6$ GeV$^2$.
The $\phi/\rho^0$ cross-section ratio is suppressed at 
lower values of $Q^2$, while
the $\omega/\rho^0$ ratio 
shows no dependence on $Q^2$.      
The $(J/\psi)/\rho^0$ ratio, on the other hand,
is a factor of two below that of the SU(4) expectation
even at $Q^2$ of $13$ GeV$^2$, although it is rising rapidly with $Q^2$.
These observations are consistent with 
a production mechanism that becomes flavour independent 
at a sufficiently high $Q^2$.

\section{Acknowledgements}

We thank the DESY Directorate for their strong support and
encouragement, and the HERA machine group for their diligent efforts.
We are grateful for the support of the DESY computing and network
services. The design, construction and installation of the ZEUS
detector have been made possible owing to the ingenuity and effort of
many people from DESY and home institutes who are not listed as
authors. It is also a pleasure to thank M. Strikman  for many
useful discussions.

\newpage

\bibliographystyle{zeusstylem}
\bibliography{zeuspubs,h1pubs,otherpubs,halina}

\begin{mcbibliography}{10}

\bibitem{rmp_50_261}
T.H. Bauer {et al.},
\newblock  Rev. Mod. Phys. { 50}  (1978)~ 261, Erratum ibid. 51 (1979)
  407\relax
\relax
\bibitem{special1}
R.M. Egloff {et al.},
\newblock  Phys. Rev. Lett. { 43}  (1979)~ 657 ;\\
R.M. Egloff {et al.},
\newblock  Phys. Rev. Lett. { 43}  (1979)~ 1545 ;\\
D. Aston {et al.},
\newblock  Nucl. Phys. { B209}  (1982)~ 56\relax
\relax
\bibitem{special2}
ZEUS Collaboration, M.~Derrick {et al.},
\newblock  Z. Phys. { C69}  (1995)~ 39 ;\\
H1 Collaboration, S.~Aid et al.,
\newblock  Nucl. Phys. { B463}  (1996)~ 3 ;\\
ZEUS Collaboration, M.~Derrick {et al.},
\newblock  Z. Phys. { C73}  (1997)~ 253\relax
\relax
\bibitem{epj_2_247}
ZEUS Collaboration, J.~Breitweg {et al.},
\newblock  Eur. Phys. J. { C2}  (1998)~ 247\relax
\relax
\bibitem{zfp_73_73}
ZEUS Collaboration, M.~Derrick {et al.},
\newblock  Z. Phys. { C73}  (1996)~ 73\relax
\relax
\bibitem{pl_377_259}
ZEUS Collaboration, M.~Derrick {et al.},
\newblock  Phys. Lett. { B377}  (1996)~ 259\relax
\relax
\bibitem{zfp_75_215}
ZEUS Collaboration, J.~Breitweg {et al.},
\newblock  Z. Phys. { C75}  (1997)~ 215\relax
\relax
\bibitem{dr_00_037}
H1 Collaboration, C.~Adloff et al.,
\newblock  DESY Report { 00-037}  (2000)~, to be published in Phys. Lett.
  B\relax
\relax
\bibitem{prl_22_981}
J.J. Sakurai,
\newblock  Phys. Rev. Lett. { 22}  (1969)~ 981\relax
\relax
\bibitem{collins}
P.D.B. Collins, {\em An Introduction to Regge Theory and High Energy Physics}
  (Cambridge University Press, 1977)\relax
\relax
\bibitem{pr_50_3134}
S.~J.~Brodsky {et al.},
\newblock  Phys. Rev. { D50}  (1994)~ 3134\relax
\relax
\bibitem{pr_54_3194}
L.~Frankfurt {et al.},
\newblock  Phys. Rev. { D54}  (1996)~ 3194\relax
\relax
\bibitem{koepf}
W. K\"opf {et al.}, in {\em Proceedings of the Workshop on Future Physics at
  HERA}, edited by G. Ingelman et al.  (DESY, Hamburg, Germany, 1996), p.~674,
  and references therein\relax
\relax
\bibitem{Collins:1997fb}
J. C. Collins, L. Frankfurt and M. Strikman,
\newblock  Phys. Rev. { D56}  (1997)~ 2982\relax
\relax
\bibitem{Ryskin:1993ui}
M. G. Ryskin,
\newblock  Z. Phys. { C57}  (1993)~ 89\relax
\relax
\bibitem{Ryskin:1995hz}
M. G. Ryskin et al.,
\newblock  Z. Phys. { C76}  (1997)~ 231\relax
\relax
\bibitem{pr_55_4329}
A.D.~Martin, M.G.~Ryskin, and T.~Teubner,
\newblock  Phys. Rev. { D55}  (1997)~ 4329\relax
\relax
\bibitem{epj_6_603}
ZEUS Collaboration, J.~Breitweg {et al.},
\newblock  Eur. Phys. J. { C6}  (1999)~ 603\relax
\relax
\bibitem{epj_13_371}
H1 Collaboration, C.~Adloff et al.,
\newblock  Eur. Phys. J. { C13}  (2000)~ 371\relax
\relax
\bibitem{pl_380_220}
ZEUS Collaboration, M.~Derrick {et al.},
\newblock  Phys. Lett. { B380}  (1996)~ 220\relax
\relax
\bibitem{dr_00_070}
H1 Collaboration, C.~Adloff et al.,
\newblock  DESY Report { 00-070}  (2000)~, submitted to Phys. Lett. B\relax
\relax
\bibitem{epj_10_373}
H1 Collaboration, C.~Adloff et al.,
\newblock  Eur. Phys. J. { C10}  (1999)~ 373\relax
\relax
\bibitem{pl_437_432}
ZEUS Collaboration, J.~Breitweg {et al.},
\newblock  Phys. Lett. { B437}  (1998)~ 432\relax
\relax
\bibitem{zeusdetector}
ZEUS Collaboration, M.~Derrick {et al.},
\newblock  {\em The ZEUS Detector},
\newblock  Status Report, 1993\relax
\relax
\bibitem{special3}
M. Derrick et al.,
\newblock  Nucl. Instr. and Meth. { A309}  (1991)~ 77 ;\\
A. Andresen et al.,
\newblock  Nucl. Instr. and Meth. { A309}  (1991)~ 101 ;\\
A. Bernstein et al.,
\newblock  Nucl. Instr. and Meth. { A336}  (1993)~ 23\relax
\relax
\bibitem{special4}
N. Harnew et al.,
\newblock  Nucl. Instr. and Meth. { A279}  (1989)~ 290 ;\\
B. Foster et al.,
\newblock  Nucl. Phys. { Proc.-Suppl. B32}  (1993)~ 181 ;\\
B. Foster et al.,
\newblock  Nucl. Instr. and Meth. { A338}  (1994)~ 254\relax
\relax
\bibitem{special5}
ZEUS Collaboration, M.~Derrick {et al.},
\newblock  Z. Phys. { C63}  (1994)~ 391 ;\\
J.~Andruszk\'{o}w {et al.},
\newblock  DESY Report { 92--066}  (1992)~\relax
\relax
\bibitem{muchorthesis}
K. Muchorowski,
\newblock  Ph.D. thesis, Warsaw University, 1998,
\newblock  (unpublished)\relax
\relax
\bibitem{heracles}
A. Kwiatkowski, H. Spiesberger and H.-J.Moehring, in {\em Proceedings of the
  Workshop on Physics at HERA}, edited by W. Buchm\"uller and G. Ingleman
  (DESY, Hamburg, Germany, 1991), p.~1294\relax
\relax
\bibitem{cpc_100_195}
M. Arneodo, L. Lamberti, and M. Ryskin,
\newblock  Comp. Phys. Comm. { 100}  (1996)~ 195\relax
\relax
\bibitem{epj_3_1}
Particle Data Group, C. Caso {et al.},
\newblock  Eur. Phys. J. { C3}  (1998)~ 1\relax
\relax
\end{mcbibliography}


\newpage

\begin{table}[htbp]
\begin{center}
\begin{small}
\begin{tabular}{|c|c|}\hline
Reaction & Cross section [nb] \\
 & $(Q^2= 7~{\gevsq}, W=70~{\gev})$ \\
\hline
$\gamma^* p \rightarrow \omega p$ & ~~$ 8.5 \pm 1.1 \pm 2.0 $ \\
\hline
$\gamma^* p \rightarrow \phi p$ & $ 17.3 \pm 5.0 \pm 6.0 $ \\
\hline
$\gamma^* p \rightarrow \rho^0 p$ & $95 \pm 9 \pm 5$\\
\hline
\hline
\multicolumn{2}{|c|}{Cross-section ratio $(Q^2= 7~{\gevsq}, W=70~{\gev})$}\\
\hline
\multicolumn{2}{|c|}{
$\sigma_{\gamma^*p \rightarrow \omega p}/
\sigma_{\gamma^*p \rightarrow \phi p} =  0.49 \pm 0.15 \pm 0.12$
}\\
\hline
\multicolumn{2}{|c|}{
$\sigma_{\gamma^*p \rightarrow \omega p}/
\sigma_{\gamma^*p \rightarrow \rho^0 p} = 
0.089 \pm 0.014 \pm 0.019$}\\
\hline
\multicolumn{2}{|c|}{
$\sigma_{\gamma^*p \rightarrow \phi p}/
\sigma_{\gamma^*p \rightarrow \rho^0 p} =
0.182 \pm 0.055 \pm 0.061$} \\
\hline
\end{tabular} 
\end{small} 
\end{center}
\caption{ 
Exclusive electroproduction cross-section measurements and ratios for 
$\omega$, $\phi$, and $\rho^0$ mesons at $Q^2 = 7$ GeV$^2$
and $W = 70$ GeV.
The $\rho^0$ cross section is taken from a previous
ZEUS publication \protect\cite{epj_6_603}.
The first uncertainty is statistical, the second systematic. 
}
\label{tab:cross} 
\end{table}
\vspace{1.5cm}
\begin{table}[htbp]
\begin{center}
\begin{small}
\begin{tabular}{|c|c|c|c|}\hline
$W$ [{\gev}] & 50 & 70 & 100 \\
\hline
Number of events & $34$ & $30$ & $59$ \\
\hline
$\sigma_{\gamma^* p \rightarrow \omega p}(Q^2=7~{\gevsq})$ [nb] &
  $6.2 \pm 1.5 \pm 1.8$ & $8.0 \pm 2.1 \pm 2.4$ &  
$12.3 \pm 2.5 \pm 3.6$ \\
\hline
\end{tabular}
\end{small}
\end{center}
\caption{
Exclusive $\omega$ production cross sections 
for $|t|<0.6$~{\gevsq} and $Q^2=7~{\gevsq}$, measured at three
$W$ values. The first uncertainty is statistical, the second systematic.
}
\label{tab:crossW}
\end{table}
\vspace{1.5cm}
\begin{table}[htbp]
\begin{center}
\begin{small}
\begin{tabular}{|c|c|c|c|}\hline
$Q^2$ [{\gevsq}] & $\simeq 0$ & 3.5 & 13 \\
\hline
$\sigma_{\gamma^* p \rightarrow \omega p}$ [nb] &
  $1210 \pm 120 \pm 230 $ & $34.2 \pm 6.8 \pm 8.6$ &  
$2.04 \pm 0.36 \pm 0.68$ \\
\hline
$\sigma_{\gamma^* p \rightarrow \rho^0 p}$ [nb] &
  $11400 \pm 300 ^{+ 1000}_{-1200} $ &
$376 \pm 27 ^{+19}_{-25}$ &  $24 \pm 3 \pm 2$ \\
\hline
$\sigma_{\gamma^* p \rightarrow \omega p}/ 
 \sigma_{\gamma^* p \rightarrow \rho^0 p} $ &
  $0.106 \pm 0.011 \pm 0.016$ & $0.091 \pm 0.019 \pm 0.020$ &  
$0.085 \pm 0.018 \pm 0.027$\ \\
\hline
\end{tabular}
\end{small}
\end{center}
\caption{
Cross-section measurements and ratios for exclusive production of
$\omega$ and $\rho^0$ mesons at three $Q^2$ values.
The ratios at $Q^2 \simeq 0~{\gevsq}$ were
calculated using photoproduction
cross sections for $\omega$ at $W=80$ GeV \protect\cite{zfp_73_73} and 
for $\rho^0$ at $W=75$ GeV~\protect\cite{epj_2_247},
whereas the cross sections at higher $Q^2$ are for $W=70$ GeV. 
The $\rho^0$ production
cross sections at high $Q^2$ are taken from a previous 
ZEUS publication~\protect\cite{epj_6_603}. 
The first uncertainty is statistical, the second systematic.
}
\label{tab:crossRatios}
\end{table}
%

%
%
%

\end{document}